\documentclass{article}
\newcommand{\cZ}{{\cal Z}}
\newcommand{\bz}{{\bar z}}

\newcommand{\hp}{{\hat p}}
\newcommand{\bhp}{{\hat{ \overline p}}}
\newcommand{\cbZ}{\overline{\cal Z}}
\newcommand{\cF}{{\cal F}}
\newcommand{\cK}{{\cal K}}
\newcommand{\cV}{{\cal V}}
\newcommand{\cbF}{\overline{\cal F}}
\newcommand{\bF}{{\overline F}{}}
\newcommand{\bxi}{{\bar\xi}}
\newcommand{\bpsi}{{\bar\psi}}
\newcommand{\I}{i}

\def\P{{\cal{M}}}         \newcommand{\Q}{{\cal{N}}}
\def\bP{\bar{\P}} \newcommand{\bQ}{\bar{\Q}}
\def\Lm{{\cal{K}}} \def\bLm{\bar\Lm}

\newcommand{\disty}{\displaystyle}
\newcommand{\ba}{\begin{array}}
\newcommand{\ea}{\end{array}}
\newcommand{\be}{\begin{equation}}
\newcommand{\ee}{\end{equation}}
\newcommand{\bea}{\begin{eqnarray}}
\newcommand{\eea}{\end{eqnarray}}
\newcommand{\bi}{\begin{itemize}}
\newcommand{\ei}{\end{itemize}}
\newcommand{\sfrac}[2]{\mbox{$\frac{#1}{#2}$}}

\newcommand{\vf}{\varphi}
\newcommand{\dvf}{\dot{\vf}}
\newcommand{\bvf}{\bar{\vf}}
\newcommand{\dbvf}{\dot{\bvf}}

\newcommand{\Vf}{\Phi}
\newcommand{\dVf}{\dot{\Vf}}
\newcommand{\bVf}{\bar{\Vf}}
\newcommand{\dbVf}{\dot{\bVf}}

\newcommand{\nn}{\nonumber}

\renewcommand{\thefootnote}{\fnsymbol{footnote}}

\usepackage{amscd,amsmath,amssymb}

\begin{document}
\thispagestyle{empty}
\begin{center}
{~}\\
\vspace{3cm}
{\Large\bf Two-dimensional N=8 supersymmetric mechanics in superspace}\\
\vspace{2cm}
{\large \bf S.~Bellucci${}^{a}$, S.~Krivonos${}^{b}$ and A.~Shcherbakov${}^{b}$ }\\
\vspace{2cm}
{\it ${}^a$INFN-Laboratori Nazionali di Frascati, C.P. 13,
00044 Frascati, Italy}\\
{\tt bellucci@lnf.infn.it} \\\vspace{0.5cm}
{\it ${}^b$ Bogoliubov  Laboratory of Theoretical Physics, JINR,
141980 Dubna, Russia}\\
{\tt krivonos, shcherb@thsun1.jinr.ru} \\ \vspace{2.0cm}

{\bf Abstract}
\end{center}

We construct a two-dimensional $N=8$ supersymmetric quantum mechanics which inherits
the most interesting properties of $N=2$, $d=4$ supersymmetric Yang-Mills theory.
After dimensional reduction to one dimension in terms of field-strength,
we show that only complex scalar fields from the $N=2, d=4$
vector multiplet become physical bosons in $d=1$.
The rest of the bosonic components are reduced to auxiliary fields,
thus giving rise to the {\bf (2,\;8,\;6)} supermultiplet
in $d=1$. We construct the most general superfields action for this supermultiplet
and  demonstrate that it possesses duality symmetry
extended to the fermionic sector of theory. We also explicitly present
the Dirac brackets for the canonical variables and construct the supercharges and Hamiltonian
which form a $N=8$ super Poincar\`{e} algebra with
central charges.
Finally,  we discuss the duality transformations which relate
the {\bf (2,\;8,\;6)} supermultiplet with the {\bf (4,\;8,\;4)} one.
\hfil
\newpage
\renewcommand{\thefootnote}{\arabic{footnote}}
\setcounter{footnote}0
\section{Introduction}
Among extended supersymmetric theories in diverse dimensions those which have
eight real supercharges are most interesting. These theories admit off-shell superfield formulations
(in the harmonic \cite{bible} or projective  \cite{pss} superspace) which greatly simplify
their analysis. Moreover, it  is possible to obtain exact quantum results for $N=2$, $d=4$ theories
in the famous Seiberg-Witten approach \cite{SW1,SW2}. Finally, the theories with eight supercharges are
the highest-$N$ case of theories with extended supersymmetries which have a rich geometric structure
of the target space (see e.g. \cite{VP1}).

One of the most investigated theories with eight supercharges is $N=2$, $d=4$ supersymmetric Yang-Mills (SYM) theory. It
has been much explored and many exciting results have been obtained. The heart of the $N=2$, $d=4$ SYM theory
is formed by a vector supermultiplet, which describes spin-1 particles, accompanied by complex scalar
fields and doublets of spinor fields. The geometry of the scalar fields is restricted to be a
K\"{a}hler one \cite{KG} of special type. The restriction that the metric be defined by a
holomorphic function is crucial for the Seiberg-Witten approach. Other interesting properties of
the $N=2$, $d=4$ SYM theory are duality in the scalar sector \cite{SW1,SW2} and the possibility of
partial spontaneous breaking of $N=2$ supersymmetry by adding two types of Fayet-Iliopoulos (FI) terms \cite{SPB,IZ}.

In \cite{Z1} it has been shown that the theories with eight supercharges can be similarly formulated
in diverse dimensions still sharing common properties. In this respect, the one-dimensional
case has a special status, because the standard reduction from the $N=2$, $d=4$ SYM to $d=1$ gives
rise to the $N=8$ supersymmetric theory with five bosons, i.e., the {\bf (5,\;8,\;3)} supermultiplet\footnote{We use the notation
$\bf (n,N,N-n)$, in order to describe a supermultiplet with $n$ physical bosons, $N$ fermions and $N-n$ auxiliary bosons.} \cite{Z1,DE}.
Of course, after such a reduction almost all nice features of $N=2$ SYM mentioned above disappear.
Naturally, an obvious question arises whether it is possible to construct an $N=8$, $d=1$ theory which
\begin{itemize}
\item contains two bosonic fields with a special K\"{a}hler
     geometry in the target-space,
\item possesses the duality transformations, properly extended to
the fermionic sector,
 \item may be obtained by reduction from the
$N=2$, $d=4$ SYM,
\item has a proper place for FI terms and therefore possesses non-trivial potentials.
\end{itemize}
In our paper \cite{BKN}  we constructed, within the Hamiltonian framework, the $N=8$ supersymmetric
mechanics (SM) which possesses the first two properties.
The goal of the present paper is to show this $N=8$ SM may be
constructed in superspace as the direct reduction from $N=2, d=4$ SYM. The main idea is to
perform the reduction to one dimension in terms of the  $N=2$ vector
multiplet $\cal A$, instead of
the reduction in terms of a prepotential \cite{Z1,DE}. In this approach
only a complex scalar from
the $N=2$, $d=4$ vector multiplet becomes a physical boson in $d=1$, while
the rest of the bosonic components
are reduced to auxiliary fields. Thus, we end up with the {\bf (2,\;8,\;6)}
supermultiplet. In Subsection \ref{2L}
we construct the most general action for this supermultiplet in terms of $N=8$ superfields with all possible FI terms included.
We also explicitly demonstrate that the action possesses duality symmetry extended to the
fermionic sector of the theory. In Subsection \ref{2H}, in order to deal with
the second-class constraints, we introduce the Dirac brackets for the canonical variables, and
find the supercharges and Hamiltonian which form $N=8$ super Poincar\`{e} algebra with
central charges. The extension of two-dimensional $N=8$ SM to the $2n$-dimensional case is performed in Subsection \ref{2nSM}.
Then, in Section \ref{484} we show that the special duality transformations relate two-dimensional $N=8$
SM with a particular case of Quaternionic SM \cite{BKSu}. Finally, in Section 4 we summarize our results and draw some conclusions.

\section{Two-dimensional N=8 SM}
In this section we describe a general superfield formalism \cite{BIKL2} and construct the most general
action for two-dimensional $N=8$ SM. We start with
the formulation of SM in  $N=8$ superspace and conclude with the component
form of the Lagrangian and Hamiltonian. The generalization
of the two-dimensional case to the $2n$-dimensional is straightforward. So, we explicitly
present here just the final results for the $2n$-dimensional SM.
A preliminary description of part of the results
was presented at a recent conference \cite{BKNS1}.

\subsection{Two-dimensional N=8 SM in superspace}\label{2L}
A convenient starting point is the $N=8$, $d=1$ superspace $\mathbb{R}^{(1|8)}$
$$
\mathbb{R}^{(1|8)}=(t,\theta^{ia},\vartheta^{i\alpha})\,,\qquad
\left(\theta^{ia}\right)^\dagger=\theta_{ia}\,,\qquad
\left(\vartheta^{i\alpha}\right)^\dagger=\vartheta_{i\alpha}\,,$$
where $i,\,a,\,\alpha=1,\,2$ are doublet indices of three $SU(2)$ subgroups of
the automorphism group of $N=8$ superspace\footnote{
We use the following convention for the skew-symmetric tensor $\epsilon$:
$\epsilon_{ij} \epsilon^{jk}=\delta_i^k$, $\epsilon_{12} = \epsilon^{21} =1.$}. In this superspace we define
the covariant spinor derivatives
\begin{equation}\label{sderiv} \ba{ll}
\disty
D^{ia}=\frac{\partial}{\partial\theta_{ia}}+\I\theta^{ia}\partial_t\,,\quad&
\disty\nabla^{i\alpha}=\frac{\partial}{\partial\vartheta_{ia}}+
\I\vartheta^{ia}\partial_t\,,\\
\disty\left\{D^{ia},D^{jb}\right\}=2\I\epsilon^{ij}\epsilon^{ab}\partial_t\,,\quad&
\disty\left\{\nabla^{i\alpha},\nabla^{j\beta}\right\}=
2\I\epsilon^{ij}\epsilon^{\alpha\beta}\partial_t\,.
\ea
\end{equation}
In full analogy with $N=2, d=4$ SYM, we introduce
a complex $N=8$ superfield $\cZ$, $\cbZ$ subjected to the following constraints:
\be \label{constr}
\ba{lc}
\displaystyle D^{1a} \cZ = \nabla^{1\alpha} \cZ =0,\qquad D^{2a} \cbZ = \nabla^{2\alpha}\cbZ=0, & \quad (a) \\
\displaystyle \nabla^{2\alpha}D^{2a} \cZ + \nabla^{1\alpha}D^{1a} \cbZ =\I M^{a\alpha} , & \quad (b)
\ea
\ee
where $M^{a\alpha}$ are arbitrary constants obeying the reality condition $\left(M^{a\alpha}\right)^\dagger=M_{a\alpha}$.
The constraints (\ref{constr}a) represent the twisted version of the standard chirality conditions, while
(\ref{constr}$b$) are recognized as modified reality constraints \cite{IZ}.
As we will see below, the presence of these arbitrary parameters $M^{a\alpha}$ gives rise to potential terms in the component action
and opens a possibility for a partial breaking of $N=8$ supersymmetry.

The constraints (\ref{constr}) leave the following components in the $N=8$ superfields $\cZ$, $\cbZ$:
\be\label{components}
\ba{llll}
z = \cZ, & \bz = \cbZ, & Y^{a\alpha}=D^{2a}\nabla^{2\alpha}\cZ, & {\overline Y}{}^{a\alpha} = -D^{1a}\nabla^{1\alpha}\cbZ=Y^{a\alpha}+iM^{a\alpha}, \\
\psi^a =D^{2a}\cZ, & \bpsi_a=-D^1_a\cbZ, & \xi^\alpha=\nabla^{2\alpha}\cZ, & \bxi_\alpha=-\nabla^1_\alpha\cbZ,\\
A=-\I D^{2a}D^2_a \cZ, & {\bar A}=-\I D^{1a}D^1_a \cbZ, & B=-\I\nabla^{2\alpha}\nabla^2_\alpha \cZ, &
{\bar B}=-\I\nabla^{1\alpha}\nabla^1_\alpha \cbZ,
\ea
\ee
where the right hand side of each expression is supposed to be taken upon $\theta^{ia}=\vartheta^{i\alpha}=0$.
The bosonic fields $A$ and $B$ are subjected, in virtue of (\ref{constr}), to the additional constraints
\be\label{constr1}
\frac{\partial}{\partial t} \left( A-{\bar B}\right)=0,\quad
\frac{\partial}{\partial t} \left( {\bar A}- B\right)=0.
\ee
In order to deal with these constraints, we have the following options:
\bi
\item to solve them as
\be\label{sol1}
A=C+\frac{m}{2},\qquad B={\overline C}-\frac{\overline m}{2},
\ee
where $C$ is a new independent complex auxiliary field and $m$ is a complex constant parameter;
the resulting supermultiplet will be just the {\bf (2,\;8,\;6)} one;
\item to insert them with Lagrange multipliers in the proper action;
this option gives rise to a {\bf (4,\;8,\;4)} supermultiplet and will be considered in section \ref{484}.
\ei
Thus, the direct reduction of the basic constraints defining the $N=2, d=4$ vector supermultiplet to
$N=8, d=1$ superspace gives (with the first option selected) the {\bf (2,\;8,\;6)} supermultiplet. Let us
stress that the possibility to
explicitly solve the constraints (\ref{constr1}) as in (\ref{sol1}) is the key feature of one-dimensional theories
which makes such a reduction to the one-dimensional supermultiplet with only two physical bosons permissible.

Now one can write down the most general $N=8$ supersymmetric Lagrangian in the $N=8$ superspace\footnote{We use
the convention $\int dt d^2\theta_2 d^2\vartheta_2\equiv \sfrac{1}{16}\int dt\; D^{2a}D^2_a\;\nabla^{2\alpha}\nabla^2_\alpha$.}
\be\label{actionsf}
\ba{ll}
\disty S=-&\disty \int dt d^2\theta_2 d^2\vartheta_2\left[ \cF\left( \cZ \right) -
  \frac{1}{2}\theta_{2a}\vartheta_{2\alpha}N^{a\alpha}\cZ -
  \frac{\I}{8}\left( {\bar n}\,\theta^a_2\theta_{2a}+ n\, \vartheta^\alpha_2\vartheta_{2\alpha}\right)\, \cZ \right]-\\[6pt]
\disty &\disty  \int dt d^2\theta_1 d^2\vartheta_1\left[ \cbF\left(\cbZ\right)+
\frac{1}{2}\theta_{1a}\vartheta_{1\alpha}N^{a\alpha}\cbZ -
 \frac{\I}{8}\left( n\, \theta_1^a\theta_{1a}+{\bar n}\, \vartheta_1^\alpha \vartheta_{1\alpha}\right)\, \cbZ \right].
\ea
\ee
Here $\cF(\cZ)$ and $\cbF(\cbZ)$ are arbitrary holomorphic functions of the superfields $\cZ$ and $\cbZ$, respectively, and the two terms
with a constant real matrix parameter $N^{a\alpha}$ ($\left( N^{a\alpha}\right)^\dagger =N_{a\alpha}$) and a complex constant
parameter $n$ represent one-dimensional versions of two FI terms \cite{IZ}.

Following the first option (\ref{sol1}), after integration over the Grassmann variables and excluding the
auxiliary fields $C, Y_{a\alpha}$ by their equations of motion, we will get the action
in terms of the physical components\footnote{All
implicit summations go from ``up-left'' to ``down-right'', e.g.,
$\psi\bpsi \equiv \psi^a\bpsi_a$, $\psi^2 \equiv \psi^a\psi_a$, $M^2\equiv M^{a\alpha}M_{a\alpha}$, etc.}
\be\label{actionc}
S=\int dt \left[ \cK -\cV \right],
\ee
where the kinetic $\cK$ and potential $\cV$ terms read
\be
\cK= \left(F''+\bF{}''\right)\left[ {\dot
z}{\dot\bz}+\sfrac{\I}{4}\left( \psi\dot\bpsi-\dot\psi\bpsi
+\xi\dot\bxi-\dot\xi\bxi\right)\right]- \sfrac{\I}{4}\left(
F'''{\dot z}-\bF{}'''\dot\bz\right)\left( \psi\bpsi+\xi\bxi\right)
\ee
and
\begin{eqnarray}
\cV&=&\frac{1}{16}\,\left(F^{(4)}-\frac{3F'''F'''}{F''+\bF{}''}\right)\psi^2\xi^2+
    \frac{1}{16}\left(\bF{}^{(4)}-\frac{3\bF{}'''\bF{}'''}{F''+\bF{}''}\right)\bpsi^2\bxi^2+\nn\\
&&    \frac{1}{16}\frac{F'''\bF{}'''}{F''+\bF{}''}\left(\psi^2\bpsi{}^2+\xi^2\bxi{}^2-4\psi\bpsi
    \xi\bxi\right)+\frac18\frac{\P\bP+\Q\bQ+\Lm\bLm}{F''+\bF''},
\eea
with
\be\label{PQ}
\P=\frac12(F'' \overline{m} - n)+iF'''\xi^2, \quad
\Q=\frac12(F'' m + \bar n)-iF'''\psi^2, \quad
\Lm_{a\alpha}=F'' M_{a\alpha} + \I N_{a\alpha}+2iF'''\psi_a\xi_\alpha.
\ee
Here the holomorphic function $F(z)$ is defined as a bosonic limit of $\cF(\cZ)$
$$ F(z)\equiv \cF(\cZ)|_{\theta=\vartheta=0}\;. $$

The action (\ref{actionc}) is invariant with respect to the $N=8$ supersymmetry which is realized on the
physical component fields as follows:
\begin{eqnarray}\label{susy1}
&&\delta z = \epsilon_{2a}\psi^a +\varepsilon_{2\alpha}\xi^\alpha ,\qquad
\delta\bz= -\epsilon_{1a}\bpsi{}^a -\varepsilon_{1\alpha}\bxi{}^\alpha,\nonumber\\
&&\delta\psi_a = \frac{\I}{2}\epsilon_{2a}\left( C+\frac{m}{2}\right)+\varepsilon^\alpha_2Y_{a\alpha}+2\I\epsilon_{1a}{\dot z},\\
&& \delta\xi_\alpha=\frac{\I}{2}\varepsilon_{2\alpha}\left( {\overline C}-\frac{\overline m}{2}\right)
-\epsilon^a_2 Y_{a\alpha}+2\I\varepsilon_{1\alpha}{\dot z},\nonumber
\\
C&=&\frac{\bP-\bQ}{F''+\bF''}, \qquad Y_{a\alpha}=-\frac{F'''\psi_a\xi_\alpha+\bF'''\bpsi_a\bxi_\alpha+\I \bLm_{a\alpha}}{F''+\bF''},
\end{eqnarray}
with $\epsilon_{ia}$, $\varepsilon_{i\alpha}$ being the parameters of two $N=4$ supersymmetries acting on
$\theta^{ia}$ and $\vartheta^{i\alpha}$, respectively.
Using the Noether theorem one can find classical expressions for the conserved supercharges
corresponding to the supersymmetry transformations (\ref{susy1})
\begin{eqnarray}\label{Q}
&& Q^a_1=\left( F''+\bF{}''\right)\psi^a\dot{\bz} -\sfrac{\I}{4}\bF{}^{(3)}\bpsi{}^a \bxi{}^2-
  \frac12 \left(\bF{}'' M^{a\alpha}-iN^{a\alpha}\right)\bxi_\alpha-
  \frac{1}{4}\left( \bF{}'' m-{\bar n}\right)\bpsi{}^a\,, \nonumber\\
&& S^\alpha_1=\left( F''+\bF{}''\right)\xi^\alpha{\dot \bz} -\sfrac{\I}{4}\bF{}^{(3)}\bxi{}^\alpha \bpsi{}^2+
  \frac12\left(\bF{}'' M^{a\alpha}-iN^{a\alpha}\right)\bpsi_a+
  \frac{1}{4}\left( \bF{}''{\overline m} +  n\right)\bxi{}^\alpha\,,\nonumber\\
&&Q_{2a}=\left( Q^a_1 \right)^\dagger,\quad S_{2a}=\left( S^a_1 \right)^\dagger.
\end{eqnarray}

Let us stress, once again, that our variant of $N=8$ SM is a reduction of
$N=2$, $d=4$ SYM.  So it is not unexpected that the metric of the bosonic manifold
is restricted to be the \emph{special K\"{a}hler} one (of rigid type)
\be\label{sk}
g(z,\bz)= F''(z)+\bF{}''(\bz).
\ee
Moreover, one may immediately check that the action  (\ref{actionc}) exhibits the
famous Seiberg-Witten duality \cite{SW1} extended to the fermionic sector of theory.
Indeed, after passing to new variables defined as
\be\label{duality}
\ba{lllllll}
\displaystyle z\rightarrow \I F'(z), &&
\displaystyle \bz\rightarrow -\I\bF'(\bz),&&
\displaystyle F''(z)\rightarrow 1/F''(z), &&
\displaystyle \bF''(\bz)\rightarrow 1/\bF''(\bz),\\[6pt]
\displaystyle \psi^a \rightarrow \I F''(z) \psi^a, &&
\displaystyle \bpsi_a \rightarrow -\I\bF''(\bz)  \bpsi_a, &&
\displaystyle \xi^\alpha \rightarrow \I F''(z) \xi^\alpha, &&
\displaystyle \bxi_\alpha \rightarrow -\I \bF''(\bz) \bxi_\alpha, \\[6pt]
\displaystyle N^{a\alpha} \rightarrow M^{a\alpha},&&
\displaystyle M^{a\alpha} \rightarrow -N^{a\alpha},&&
\displaystyle m \rightarrow \bar n,&&
\displaystyle n \rightarrow {\overline m},\\[6pt]
\ea
\ee
the action (\ref{actionc}) keeps its form.
Let us note that in the dual formulation the constants
$M^{a\alpha}$ and $m$, which appear in the constraints (\ref{constr}) and (\ref{sol1}),
are interchanged with the constants $N^{a\alpha}$ and $n$, which have
shown up in the FI terms.
This is just a simplified version of the electric-magnetic duality \cite{SW1} for our
$N=8$ SM case. Thus, our $N=8$ SM possesses the most interesting peculiarities
of the $N=2$, $d=4$ SYM theory and can be used for a simplified analysis of some
subtle properties of its ancestor.

\subsection{Two-dimensional N=8 SM: Hamiltonian}\label{2H}
In order to find the classical Hamiltonian, we follow the standard procedure of
quantizing a system with bosonic and fermionic degrees of freedom.
{}From the action (\ref{actionc}) we  define the momenta $p$, ${\bar p}$, $\pi^{(\psi)}_a$,
$\bar\pi^{(\psi)a}$, $\pi^{(\xi)}_\alpha$, $\bar\pi{}^{(\xi)\alpha}$ conjugated to
$z$, $\bz$, $\psi^a$, $\bpsi_a$, $\xi^\alpha$ and $\bxi_\alpha$, respectively, as
\be\label{momenta}
\ba{lc}
\disty p=g\dot\bz-\frac{\I}{4}\, \partial_z g \left(\psi\bpsi+\xi\bxi\right),\quad
\bar p= g{\dot z}+\frac{\I}{4}\, \bar\partial_z g \left(\psi\bpsi+\xi\bxi\right),& (a) \\[1.5mm]
\disty \pi^{(\psi)}_a=-\frac{\I}{4}\, g\bpsi_a,~
\bar\pi{}^{(\psi)a}=-\frac{\I}{4}\, g \psi^a, \quad
\pi^{(\xi)}_\alpha=-\frac{\I}{4}\,g\bxi_\alpha,~
\bar\pi{}^{(\xi)\alpha}=-\frac{\I}{4}g\, \xi^\alpha, &(b)
\ea
\ee
with the metric $g(z,\bz)$ defined in (\ref{sk})
and introduce Dirac
brackets for the canonical variables
\be\label{pb2}
\ba{lll}
\disty\left\{ z,\,\hp \right\}=1, &
\disty\left\{ \bz, \bhp\right\}=1, &
\disty\left\{ \hp,\,\bhp\right\}=-\frac{\I}{2}\frac{\partial_z g\, \bar\partial_z g }{g}\left( \psi\bpsi +\xi\bxi \right),\\[6pt]
\disty\left\{ \hp,\,\psi_a \right\}=\frac{\partial_z g}{g}\,\psi_a, &
\disty\left\{ \bhp,\,\bpsi_a \right\}=\frac{\bar\partial_z g}{g}\,\bpsi_a,&
\disty\left\{ \psi^a,\,\bpsi_b \right\}=-\frac{2\I}{g}\,\delta^a_b,\\[6pt]
\disty\left\{ \hp,\,\xi_\alpha \right\}=\frac{\partial_z g}{g}\,\xi_\alpha,&
\disty\left\{ \bhp,\,\bxi_\alpha \right\}=\frac{\bar\partial_z g}{g}\,\bxi_\alpha,&
\disty\left\{ \xi^\alpha,\,\bxi_\beta \right\}=-\frac{2\I}{g}\,\delta^\alpha_\beta,
\ea
\ee
where the ``improved'' bosonic momenta have been defined as
\be
\hp \equiv p+\frac{\I}{4}\, \partial_z g\left(\psi\bpsi +\xi\bxi \right),\quad
\bhp \equiv \bar p-\frac{\I}{4}\, \bar\partial_z g\left(\psi\bpsi +\xi\bxi \right).
\ee
Now one can check that the supercharges $Q_{ia}$, $S_{i\alpha}$ (\ref{Q}), being rewritten through
the momenta as
\bea\label{nQ}
 Q^a_1&=&\hp\,\psi^a -\frac{\I}{4}\,\bar\partial_z g\, \bpsi{}^a \bxi{}^2-
 \frac12 \left(\bF{}'' M^{a\alpha}-iN^{a\alpha}\right)\bxi_\alpha-
  \frac{1}{4}\left( \bF{}'' m-{\bar n}\right)\bpsi{}^a\,,\nonumber\\
 S^\alpha_1&=&\hp\,\xi^\alpha-\frac{\I}{4}\,\bar\partial_z g \, \bxi{}^\alpha \bpsi{}^2+
 \frac12\left(\bF{}'' M^{a\alpha}-iN^{a\alpha}\right)\bpsi_a+
  \frac{1}{4}\left( \bF{}'' {\overline m} + n\right)\bxi{}^\alpha\,, \\
 Q_{2a}&=&\left( Q^a_1 \right)^\dagger\,,\quad S_{2a}=\left( S^a_1 \right)^\dagger\,,\nonumber
\eea
and the Hamiltonian
\be\label{ham}
H=g^{-1}\, \hp \bhp +{\cal V}
\ee
form the  $N=8$ Poincar\`e superalgebra with the central charges
\bea\label{SA}
\left\{ Q_{ia},\,Q_{jb}\right\}&=&-2\I
\epsilon_{ij}\epsilon_{ab}\left( H-\sfrac{1}{16}\left( nm+{\bar
n}{\overline m}\right)\right)-
 \sfrac{1}{8}\,\epsilon_{ij}\left( N_a^\alpha M_{\alpha b}+
 N_b^\alpha M_{\alpha a}\right), \nonumber\\
\left\{ S_{i\alpha},\,S_{j\beta}\right\}&=&-2\I
\epsilon_{ij}\epsilon_{\alpha\beta}\left( H+\sfrac{1}{16}\left(
nm+{\bar n}{\overline m}\right)\right)-
 \sfrac{1}{8}\,\epsilon_{ij}\left( N_\alpha^a M_{a \beta}+
 N_\beta^a M_{a\alpha}\right), \nonumber\\
\left\{ Q_{1a},\,S_{2\alpha}\right\}&=&-m N_{a\alpha}-\I{\bar
n}M_{a\alpha}\,,\quad
\left\{Q_{2a},\,S_{1\alpha}\right\}=-{\overline m} N_{a\alpha}+\I n
M_{a\alpha}\,.
\eea

By these we complete the classical description of the two-dimensional $N=8$ SM.

Before closing this Subsection and going on to generalize our SM to the $2n$-dimensional case,
let us briefly summarize the main peculiarities of the model.

Firstly, as it has already been mentioned, the $N=8$ supersymmetry strictly fixes the metric of the target space to be
the \emph{special K\"{a}hler} one.

Next, the presence of the central charges in the superalgebra (\ref{SA}), as in the $N=4$ SM case \cite{IKP},
is the most exciting feature of the model. The central charges appear only when the FI terms are added
(with the constants $N_{a\alpha}$ or $n$) and the
auxiliary fields contain the constant parts ($M_{a\alpha}$ or $m$). The existence of the nonzero central charges in the
superalgebra (\ref{SA}) opens up the possibility of realizing a partial spontaneous breaking of $N=8$ supersymmetry.

Finally, it is worth noticing that the bosonic potential terms, which appear in the Hamiltonian, explicitly break
at least one of the $SU(2)$ automorphism groups. This feature is, once again, very similar to the case of $N=4$ SM \cite{IKP}.

\subsection{2n-dimensional N=8 SM}\label{2nSM}
The generalization of the $N=8$ two-dimensional SM to the $2n$-dimensional case is straightforward.
The simplest one is the superfield generalization. The related steps are described in the following.
\bi
\item We introduce $n$ complex $N=8$ superfields $\cZ^{A}$, $\cbZ{}^{B}$
$(A,B=1,\ldots,n)$, each of them obeys the
same constraints (\ref{constr}) with different constants $M^{A\,a\alpha}$
\be\label{2nconstr}
\ba{lc}
\displaystyle  D^{1a} \cZ^{A} = \nabla^{1\alpha} \cZ^{A} =0,\qquad D^{2a} \cbZ{}^{A} = \nabla^{2\alpha}\cbZ{}^{A}=0, & \quad (a) \\[1mm]
\displaystyle  \nabla^{2\alpha}D^{2a} \cZ^{A} + \nabla^{1\alpha}D^{1a} \cbZ{}^{A} =\I M^{A\, a\alpha}. & \quad (b)
\ea
\ee
\item The components of each superfield can be defined as in (\ref{components}) and $n$ different constants $m^{A}$ may
be introduced similarly to (\ref{sol1})
\be\label{2nsol1}
A^{A}=C^{A}+\frac{m^{A}}{2},\qquad B^{A}={\overline C}{}^{A}-\frac{{\bar m}^{A}}{2}.
\ee
\item The most general $N=8$ supersymmetric action reads
\begin{eqnarray}\label{actionsf2n}
S_{2n}&=&-\int dt d^2\theta_2 d^2\vartheta_2\left[ \cF(\cZ^{1},\ldots,\cZ^{k}) -
  \frac{1}{2}\theta_{2a}\vartheta_{2\alpha}\sum_A N^{a\alpha}_{A}\cZ^{A} - \right.\nonumber\\
&& \left. \frac{\I}{8}\sum_A \left( {\bar n}_{A}\,\theta^a_2\theta_{2a}+ n_{A}\,
\vartheta^\alpha_2\vartheta_{2\alpha}\right)\, \cZ^{A} \right]
+c.c.,
\end{eqnarray}
where $\cF(\cZ^{1},\ldots,\cZ^{k})$, $\cbF(\cbZ{}^{1},\ldots,\cbZ{}^{k})$ are arbitrary holomorphic functions of
the $n$ superfields $\cZ^{A}$ and $\cbZ{}^{A}$,
respectively, and all possible FI terms with the constants $N^{a\alpha}_{A}$ and $n_{A}$ are included.
\ei
The rest of the calculations goes in the same way as it is done in the previous subsections. For completeness,
we present here the explicit structure of the Dirac brackets between the canonical variables
\begin{eqnarray}\label{2nPB}
&&\left\{ z^A,\,\hp_B \right\}=\delta^A_B,\quad
\left\{ \bz{}^A,\, \bhp_B\right\}=\delta^A_B,\nonumber\\
&&
\left\{ \hp_A,\,\bhp_B\right\}=-\frac{\I}{2}g^{EE'}\partial^3_{ACE} F\,\bar\partial^3_{BC'E'}\bF
\left( \psi^{a\,C}\bpsi^{C'}_a +\xi^{\alpha\, C}\bxi^{C'}_\alpha \right), \nonumber\\
&&
\left\{ \psi^{Aa},\,\bpsi^B_b \right\}=- 2\I g^{AB} \delta^a_b,\quad
\left\{ \xi^{A\alpha},\,\bxi^B_\beta \right\}=- 2\I g^{AB}\delta^\alpha_\beta, \nonumber\\
&& \left\{ \hp_A,\,\psi^B_a \right\}=g^{BC}\partial^3_{ACE}F\,\psi^E_a,\quad
\left\{ \hp_A,\,\xi^B_\alpha \right\}=g^{BC}\partial^3_{ACE}F\,\xi^E_\alpha, \nonumber\\
&& \left\{ \bhp_A,\,\bpsi^B_a \right\}=g^{BC}\bar\partial^3_{ACE}\bF\, \bpsi^E_a,\quad
\left\{ \bhp_A,\,\bxi^B_\alpha \right\}=g^{BC}\bar\partial^3_{ACE}\bF\, \bxi^E_\alpha,
\end{eqnarray}
where the metric $g_{AB}$ and its inverse $g^{AB}$ are defined as
\be\label{2nmetric}
g_{AB}=\frac{\partial^2}{\partial z^A \partial z^B} F(z^1,\ldots,z^k)+
 \frac{\partial^2}{\partial \bz^A \partial \bz^B} \bF (\bz^1,\ldots,\bz^k),\qquad
g^{AB}g_{BC}=\delta^A_C.
\ee
Like to the previous section it is convenient to introduce $n$-dimensional extension of the quantities
(\ref{PQ})
\be
\ba{ll}
\P_A=\frac12(\partial^2_{AB}F {\overline m}^B - n_A)+\I \partial^3_{ABC}F\, \xi^{B\alpha}\xi^C_\alpha,\quad &
\Q_A=\frac12(\partial^2_{AB}F  m^B + \bar n_A)-\I \partial^3_{ABC}F\, \psi^{B a}\psi^C_a,\\ [6pt]
\Lm_A^{a\alpha}=\partial^2_{AB}F M^{B a\alpha} +\I N^{a\alpha}_A+2i\partial^3_{ABC}F\,\psi^{B a}\xi^{C\alpha}&
\ea
\ee
to have the supercharges and Hamiltonian been written in term of these objects quite elegantly.
The supercharges
\bea\label{2nSC}
Q_1^a&=&\hp_A\psi^{Aa}
    -\frac {\I}4 \bar\partial_{ABC}^3\bF\bpsi^{Aa}\bxi^{B\alpha}\bxi^C_\alpha
    -\frac12 (\bar\partial^2_{AB}\bF M^{Ba\alpha}-\I N_A^{a\alpha})\bxi^A_\alpha
    -\frac14 (\bar\partial^2_{AB}\bF m^B-\bar n_A ) \bpsi^{Aa},\nn\\
S_1^\alpha&=&\hp_A\xi^{A\alpha}
    -\frac {\I}4 \bar\partial_{ABC}^3\bF\bxi^{A\alpha}\bpsi^{Ba}\bpsi^C_a
    +\frac12(\bar\partial^2_{AB}\bF M^{Ba\alpha}-\I N_A^{a\alpha}) \bpsi^A_a
    +\frac14 (\bar\partial^2_{AB}\bF{\overline m}^B+n_A ) \bxi^{A\alpha}, \\
Q_{2a}&=&\left( Q^a_1 \right)^\dagger\,,\qquad S_{2a}=\left( S^a_1 \right)^\dagger\nn
\eea
and the Hamiltonian
\begin{eqnarray}\label{H2n}
H_{2n}&=& g^{AB}\hp_A \bhp_B+ \nonumber\\
&&\frac{1}{16}\left( \partial^4_{ABCD} F
-g^{EE'}\partial^3_{ABE}F\,\partial^3_{CDE'}F-
    2g^{EE'}\partial^3_{ACE}F\,\partial^3_{BDE'}F\right)\psi^{Aa}\psi^B_a\xi^{C\alpha}\xi^D_\alpha+\nonumber\\
&&\frac{1}{16}\left( \bar\partial^4_{ABCD} \bF
-g^{EE'}\bar\partial^3_{ABE}\,\bF\bar\partial^3_{CDE'}\bF-
    2g^{EE'}\bar\partial^3_{ACE}\bF\,\bar\partial^3_{BDE'}\bF\right)\bpsi^{Aa}\bpsi^B_a\bxi^{C\alpha}\bxi^D_\alpha+\nonumber\\
&&\frac{1}{16}\,g^{EE'}\partial^3_{ABE}F\bar\partial^3_{CDE'}\bF\left(
\psi^{Aa}\psi^B_a \bpsi^{Cb}\bpsi^D_b+
    \xi^{A\alpha}\xi^B_\alpha \bxi^{C\beta}\bxi^D_\beta -4 \psi^{Aa}\bpsi^C_a \xi^{B\alpha}\bxi^D_\alpha\right)+\nonumber\\
&&\frac18\,g^{AB}\left[ \P_A\bP_B + \Q_A\bQ_B + \Lm^{a\alpha}_A \bLm_{B\,a\alpha} \right]
\end{eqnarray}
form the  superalgebra
$$
\left\{ Q_{ia},\,Q_{jb}\right\}=-2\I
\epsilon_{ij}\epsilon_{ab}\left( H-\sfrac1{16} (n_A m^A+\bar n_A {\overline m}^A)\right)-
 \sfrac{1}{8}\,\epsilon_{ij}\left( N_{A\,a}^\alpha M^A_{\alpha b}+
 N_{A\, b}^\alpha M^A_{\alpha a}\right),
$$
$$
\left\{ S_{i\alpha},\,S_{j\beta}\right\}=-2\I
\epsilon_{ij}\epsilon_{\alpha\beta}\left( H+\sfrac{1}{16}\left(
n_A m^A+{\bar n}_A{\overline m}^A\right)\right)-
 \sfrac{1}{8}\,\epsilon_{ij}\left( N_{A\,\alpha}^a M^A_{a \beta}+
 N_{A\,\beta}^a M^A_{a\alpha}\right),
 $$
\be\label{SA1}
\left\{ Q_{1a},\,S_{2\alpha}\right\}=-m^A N_{A\,a\alpha}-\I{\bar
n}_A M^A_{a\alpha}\,,\quad
\left\{Q_{2a},\,S_{1\alpha}\right\}=-{\overline m}^A N_{A\,a\alpha}+\I n_A
M^A_{a\alpha}\,.
\ee

\section{{\bf (4,\,8,\,4)} supermultiplet}\label{484}
In this Section  we discuss the duality transformations which relate
the {\bf (2,\;8,\;6)} supermultiplet with the {\bf (4,\;8,\;4)} one.

As it has already been mentioned in Subsection \ref{2L} there is the other possibility to deal with the constraints
(\ref{constr1}). Namely, one can incorporate them into the action using the Lagrange multipliers $\vf$, $\bvf$
\bea\label{action1}
S&=&\int dt \left\{ \left(F''+\bF{}''\right)\left[ {\dot z}{\dot\bz}+\frac{i}{4}\left(
\psi\dot\bpsi-\dot\psi\bpsi +\xi\dot\bxi-\dot\xi\bxi\right)\right]-
\frac{i}{4}\left( F^{(3)}{\dot z}-\bF{}^{(3)}\dot\bz\right)\left( \psi\bpsi+\xi\bxi\right)+\right.\nn\\
&& \frac{1}{16}\left( F''\left(2Y^2+AB\right) +\bF{}''\left( 2Y^2+{\bar A}{\bar B}\right)-
F^{(4)}\psi^2\xi^2 -\bF{}^{(4)}\bpsi{}^2\bxi{}^2 \right. - \nn\\
&&\left. F^{(3)}\left( i A\xi^2+i B \psi^2 -4\psi^a \xi^\alpha Y_{a\alpha}\right) -
\bF{}^{(3)}\left(i{\bar A}\bxi{}^2 +i {\bar B} \bpsi{}^2 +4 \bpsi{}^a\bxi{}^\alpha Y_{a\alpha}\right)\right)+\nn\\
&& \left. \frac14 \varphi \left(\dot A-\dot{\bar B}\right) +
    \frac14 \bar\varphi \left( \dot{\bar A}-\dot B\right)\right\} .
\eea
For the sake of simplicity we consider the action (\ref{actionsf}) with all constant parameters equal to zero.
Eliminating the auxiliary fields $A$, $B$ and $Y_{a\alpha}$ via their equations of motion
\be\label{aux}
A=\frac{iF^{(3)}\psi^2-4\dbvf}{F''},\qquad
B=\frac{iF^{(3)}\xi^2+4\dvf}{F''},\qquad
Y_{a\alpha}=\frac{\bF{}^{(3)}\bpsi_a\bxi_\alpha -F^{(3)}\psi_a\xi_\alpha}
            {F''+\bF{}''}
\ee
one gets
the following expressions for the kinetic
\bea\label{kin}
\cK&=&\left(F''+\bF{}''\right)\left[ {\dot z}{\dot\bz}+\frac{i}{4}\left(
\psi\dot\bpsi-\dot\psi\bpsi +\xi\dot\bxi-\dot\xi\bxi\right)\right]+
\left(\frac1{F''}+\frac1{\bF''}\right)\dvf\dbvf-\nn\\
&-&\frac{i}{4}\left( F'''{\dot z}-\bF{}'''\dot\bz\right)
\left( \psi\bpsi+\xi\bxi\right)-
\frac i4 \dvf\left(\frac{F'''}{F''}\psi^2 -\frac{\bF'''}{\bF''}\bxi^2\right)-
\frac i4 \dbvf\left(\frac{\bF'''}{\bF''}\bpsi^2 -\frac{F'''}{F''}\xi^2\right)
\eea
and potential
\bea\label{pot}
\cV&=& \frac{1}{16}\left( F^{(4)} -\frac{3F'''F'''}{F''+\bF{}''}-
\frac{F'''^2\bF''}{(F''+\bF'')F''} \right)\psi^2\xi^2 +\nn\\
&&+\frac1{16}\left( \bF{}^{(4)} -\frac{3\bF{}'''\bF{}'''}{F''+\bF''}-
\frac{\bF'''^2 F''}{(F''+\bF'')\bF''} \right)\bpsi^2\bxi^2 +
\frac{F'''\bF'''}{4(F''+\bF'')}\psi\bpsi \xi\bxi
\eea
terms of the action (\ref{action1}).
{}From Eq.(\ref{kin}) one sees that the field $\vf$ becomes dynamical, giving rise to SM  with
four physical bosons.

The possibility to invert the auxiliary fields to the physical ones in one dimension was
noticed many years ago in \cite{GR}. The simplest way to demonstrate this is to
consider the transformation properties of the auxiliary fields $A$ and $B$ which read
\be\label{trAB}
\delta A = 4\varepsilon_2^\alpha \dot{\bar{\xi}}_\alpha+4\epsilon_1^a \dot{\psi}_a, \qquad
\delta B = 4\epsilon_2^a \dot{\bar{\psi}}_a + 4\varepsilon_1^\alpha \dot{\xi}_\alpha \;.
\ee
Due to the fact that the r.h.s. of (\ref{trAB}) is a total time derivative,
one can make the auxiliary fields ``dynamical'' by setting
\be\label{sub}  A=\bar B = 4\dVf,\qquad \bar A = B = 4\dbVf \;.
\ee
The new dynamical fields $\Phi$ and $\bar{\Phi}$ transform under $N=8$ supersymmetry as
\be\label{trABa}
\delta \Phi = \varepsilon_2^\alpha {\bar{\xi}}_\alpha+\epsilon_1^a {\psi}_a, \qquad
\delta \bar{\Phi} = \epsilon_2^a {\bar{\psi}}_a + \varepsilon_1^\alpha {\xi}_\alpha
\ee
and form, with the remaining components from (\ref{components}), the ({\bf 4,8,4}) supermultiplet.
Substituting (\ref{sub}) into the action (\ref{action1}) one can get
\bea\label{kin2}
\cK&=& \left(F''+\bF{}''\right)\left[ {\dot z}{\dot\bz}+\dVf\dbVf+
\frac{i}{4}\left(
\psi\dot\bpsi-\dot\psi\bpsi +\xi\dot\bxi-\dot\xi\bxi\right)\right]-\nn\\
&-&\frac{i}{4}\left( F'''{\dot z}-\bF{}'''\dot\bz\right)
\left( \psi\bpsi+\xi\bxi\right)-
\frac i4 \dVf\left(F'''\xi^2 +\bF'''\bpsi^2\right)-
\frac i4 \dbVf\left(F'''\psi^2 +\bF'''\bxi^2\right),
\\
\label{pot2}
\cV&=& \frac{1}{16}
\left( F^{(4)} -\frac{2F'''F'''}{F''+\bF{}''}\right)\psi^2\xi^2 +
\frac1{16}\left( \bF{}^{(4)} -\frac{2\bF{}'''\bF{}'''}{F''+\bF''}\right)\bpsi^2\bxi^2 +
\frac{F'''\bF'''}{4(F''+\bF'')}\psi\bpsi \xi\bxi.
\eea
An amazing fact is that the actions corresponding to Eqs.(\ref{kin}), (\ref{pot}) and Eqs.(\ref{kin2}), (\ref{pot2}) are transformed
into each other by the same Seiberg-Witten duality transformations (\ref{duality})
(with the constants $M$, $N$, $m$ and $n$ omitted) augmented by
\be\label{dual}
\vf \rightarrow \I \bVf, \quad
\bvf \rightarrow -\I\Vf, \quad
\Vf \rightarrow \I\bvf, \quad
\bVf \rightarrow -\I\vf.
\ee
Thus we conclude, that two-dimensional $N=8$ SM is dual to $N=8$ Quaternionic SM \cite{BKSu}
with the additional restriction on the target space metric to be a special K\"{a}hler one depending only
on two fields $z,\bar z$.

\section{Summary and conclusions}

In the present paper we give a superfield description  of $N=8$ SM with {\bf (2,\;8,\;6)}
supermultiplet. This supermultiplet is obtained by a direct reduction from the
$N=2$, $d=4$ vector supermultiplet. We constructed the
most general action in $N=8$ superspace with all possible FI terms and explicitly showed that
the geometry of the target space is restricted to be the special K\"{a}hler one.
Apart from the
$N=8$ superfield formulation, we presented the component action with all auxiliary fields,
as well as with the physical fields only. As a nice feature, the constructed action possesses a
duality which acts not only in the bosonic sector, but also in the fermionic one.
We performed the Hamiltonian analysis and found the
Dirac brackets between the canonical variables. The supercharges and Hamiltonian form
a $N=8$ super Poincar\`{e} algebra with central charges. The latter are proportional
to the product of two constants --- one that comes from the FI terms, and the other that appears in the
superfield constraints. These constants are directly related to the appearance of potential terms
in the Hamiltonian. We also presented the extension of the $N=8$ two-dimensional SM to the
$2n$-dimensional case. The possibility to revert auxiliary fields to physical ones has been used, in order
to construct a dual version of $N=8$ SM, which is just a special case of $N=8$ Quaternionic SM.

These results should be regarded as preparatory for a more detailed study
of $2n$-dimensional SM
with $N=8$ supersymmetry. In particular, it would be interesting to construct
the full quantum version
with some specific K\"{a}hler potential. Generally speaking, we believe that
just this version
of SM could be rather useful for a simplified analysis of subtle problems
which appear in the $N=2$, $d=4$ SYM.
For example, one may try to fully analyse the effects of
non-anti-commutativity in superspace \cite{nac},
including modifications of the spectra, etc.

Finally, due to the appearance of the central charges in the $N=8$ Poincar\`{e} superalgebra one may expect
the existence of different patterns of partial supersymmetry breaking, like in the $N=4$ SM case \cite{IKP,P1}.

\section*{Acknowledgements}
We would like to thank A.~Nersessian for many useful discussions and for the collaboration on the
earlier stage of this investigation.

This research was partially supported by the European Community's Marie Curie Research Training Network
under contract MRTN-CT-2004-005104 Forces Universe,
INTAS-00-00254 grant,
RFBR-DFG grant No 02-02-04002, grant DFG No 436 RUS 113/669, and RFBR grant
No 03-02-17440.
S.K. thanks INFN --- Laboratori Nazionali di Frascati  for the warm
hospitality extended to him during the course of this work.

\end{document}